# Toward a Wearable RFID System for Real-Time Activity Recognition Using Radio Patterns

Liang Wang, *Member, IEEE*, Tao Gu, *Senior Member, IEEE*, Xianping Tao, *Member, IEEE*, and Jian Lu, *Member, IEEE*

**Abstract**—Elderly care is one of the many applications supported by real-time activity recognition systems. Traditional approaches use cameras, body sensor networks, or radio patterns from various sources for activity recognition. However, these approaches are limited due to ease-of-use, coverage, or privacy preserving issues. In this paper, we present a novel wearable Radio Frequency Identification (RFID) system aims at providing an easy-to-use solution with high detection coverage. Our system uses passive tags which are maintenance-free and can be embedded into the clothes to reduce the wearing and maintenance efforts. A small RFID reader is also worn on the user's body to extend the detection coverage as the user moves. We exploit RFID radio patterns and extract both spatial and temporal features to characterize various activities. We also address the issues of false negative of tag readings and tag/antenna calibration, and design a fast online recognition system. Antenna and tag selection is done automatically to explore the minimum number of devices required to achieve target accuracy. We develop a prototype system which consists of a wearable RFID system and a smartphone to demonstrate the working principles, and conduct experimental studies with four subjects over two weeks. The results show that our system achieves a high recognition accuracy of 93.6 percent with a latency of 5 seconds. Additionally, we show that the system only requires two antennas and four tagged body parts to achieve a high recognition accuracy of 85 percent.

## 1 INTRODUCTION

As an enabling technology, real-time human activity recognition plays a central role in many applications. One of the critical applications attracts much research interest is elderly care because of the growing number of elderly people around the world [1]. Studies show that aged persons experience steady decline in cognitive, visual and physical functions cased by different age-related diseases [2]. New applications are under active development to provide daily support for elderlies with different types and degrees of impairments. For example, smart reminder systems [1] monitor the user's activities and reminds her when planned daily routines are not followed. Emergency response systems [3] send alarms on detecting dangerous user behaviors such as falling. Long-term health monitoring systems [4] monitor the user's activities over time for health assessment and monitoring.

Building an activity recognition system for the elderly users with possible cognitive and physical impairments poses several challenges. *Ease-of-use*—Most systems are designed for personal usage scenarios which the user receives minimum supervision from the professionals [4]. As a result, it should require minimum effort to configure and maintain the system, especially for elderlies with cognitive and physical impairments. *Coverage*—The system's working area should not be limited considering the mobility of an user. Tracking elderlies mobility and activity has shown be a vital evidence for determining their quality of life [5]. *Privacy Preserving*—The system should not be invasive to user privacy [6], especially for elderly users under long-term monitoring.

Due to the fact that computer vision-based approaches are usually privacy invasive [7], wearable sensor-based approaches have been widely adopted for activity recognition [7], [8], [9], [10], [11], [12]. In this paradigm, a smartphone and/or multiple sensor nodes are worn on an user's body to capture rich sensor data related to her/his activities. The sensor nodes form a body sensor network (BSN) to capture, deliver and process motion data to recognize user's activities. Existing work has shown the effectiveness of wearable sensors for activity recognition with fine-grained and high-coverage recognition abilities. However, they may not apply to elderly users for ease-of-use problems like sensitivity to sensor displacement [4] and requiring constant battery maintenance [13].

To address the above limitations, recent work explores device-free activity recognition approaches using RF signals [14], [15], [16], [17], [18], [19], [20], [21]. By correlating RF signal fingerprints with different activities, this work performs activity recognition using RF signal features extracted from Radio Signal Strength (RSS) [14], [20] or Channel State Information (CSI) [17], [19]. Compared to BSN-based approaches, the advantage of the above approaches lies in that they pose no or minimum requirements on the user to wear devices for activity recognition. The downside of these approaches is that they require the transmission/receiving

- *L. Wang, X. Tao, and J. Lu are with the State Key Laboratory of Novel Computer Software, Nanjing University, Nanjing, Jiangsu, P.R. China. E-mail: {wl, txp, lj}@nju.edu.cn.*
- *T. Gu is with the School of Computer Science and Information Technology, RMIT University, Melbourne, Vic., Australia. E-mail: tao.gu@rmit.edu.au.*

*Manuscript received 22 Apr. 2015; revised 22 Feb. 2016; accepted 23 Feb. 2016. Date of publication 3 Mar. 2016; date of current version 1 Dec. 2016.*
*For information on obtaining reprints of this article, please send e-mail to: reprints@ieee.org, and reference the Digital Object Identifier below.*
*Digital Object Identifier no. 10.1109/TMC.2016.2538230*

devices to be located near the user, limiting the coverage area of the detection system with respect to user's mobilities. Though much work has focused on activity recognition in home settings [22], studies also point out the importance to provide wide detection coverage involving user's outdoor activities [23].

To provide an easy-to-use activity recognition system with high coverage, this paper proposes a novel approach that combines the benefits of both the wearable sensor-based and the RF-based systems. Similar to RF-based approaches [15], the proposed system uses RF signals from wearable passive RFID tags for activity sensing. The passive tags are low-cost, battery-free, encapsulated devices that can be embedded to user's clothes. Once embedded, these passive tags require no further wearing and maintenance efforts as the BSN nodes and can survive in harsh conditions such as laundry. Different from existing RF-based approaches, our system also places a small UHF RFID reader with its antennas on the user's body. With this only active sensing device used, we greatly extend the coverage of our system compared to existing RF-based solutions while keeping the wearing and maintenance efforts to the minimum. Existing work and our previous experience in RF-based activity recognition [18], [20] suggest the effectiveness of using RSS of wireless links for activity recognition. As a result, in this work, we use Commercial Off-the-Shelf (COTS) RFID readers with RSS sensing ability instead of more sophisticated devices with CSI and signal phase capturing abilities to provide a lower-cost and more accessible solution. Recent technical trends show that low-cost, low-power RFID readers are becoming commonly available on smartphones [24], making it promising for our work to be widely adopted in the future.

The feasibility of our wearable RFID system for activity recognition is supported by two observations: 1) there exists heavy attenuation of the human body to radio communication band in which the UHF RFID operates, and 2) RFID radio communication is highly affected by the tag-antenna distance and orientation. Based on these observations, if we deploy an RFID system on a human body, user motions will result in different radio patterns which can differentiate activities. In our system, channel reliability and fading provide useful information for activity recognition. Preliminary studies presented in Section 3 show the potential to discriminate different activities by RSS information obtained from our wearable RFID system.

To support practical applications and perform reliable activity recognition using the proposed wearable RFID system, the following challenges must be addressed. First, false negative of tag readings and tag/antenna calibration are two major challenges in our system. Existing work proposes solutions such as using reference tags and conducting preliminary experiments [14], [25]. However, these solutions are not applicable in our scenario because we aim at building a mobile recognition system which works at open, diverse environments. In this work, we propose an efficient data completion technology to address the false negative reading problem. Temporal and spatial features are extracted and carefully selected to characterize the radio patterns while tolerate large variances in tag performance. Second, the number of tags and antennas should be minimized to reduce cost and improve user comfort. In this paper, we present a wrapper algorithm for antenna and tag selection which minimizes the number of devices needed with respect to the recognition accuracy requirements. Finally, to support applications with real-time requirements [1], [3], we develop a framework for real-time activity recognition on the smartphones. We use a fixed-length sliding window to control latency bound, and develop a fast, lightweight, online algorithm based on Support Vector Machine (SVM) to be executed on smartphones. We conduct comprehensive experiments involving multiple human subjects. The results show that our system is capable to conduct accurate and real-time activity recognition.

To sum up, this paper makes the following contributions.

- We propose a novel wearable activity sensing architecture involving both the passive RFID-tags and the reader which provides an easy-to-use system with high coverage suitable for applications such as elderly care.
- We implement the prototype system and carry out preliminary studies to demonstrate the feasibility of the proposed system for activity recognition.
- The feature extraction, data completion, antenna and tag selection, and real-time activity recognition algorithms are proposed to address the practical challenges with wearable RFID-based activity recognition.
- Extensive experiments and comparison studies are carried out to evaluate the performance of the proposed approach in diverse aspects.

The rest of the paper is organized as follows. Section 2 introduces related work. Hardware setup and preliminary experiment results are presented in Section 3. We present the details of our system design in Section 4. Section 5 reports results of empirical studies and Section 6 concludes the paper.

This paper is an extended version of our previous work [26]. The key contributions of this work include new algorithm design and experiments concerning the practical issues in real usage scenarios. More specifically, first, we present a wrapper algorithm for antenna and tag selection that minimizes the amount of antennas and tags required regarding an accuracy threshold. Second, we conduct extensive experiments on how training data quality affects system performance in multiple aspects and propose our solutions to address data dependency issues. Third, we compare our system with other wearable sensor-based systems in terms of recognition accuracy, battery, and total cost. Moreover, we give the details for the data completion algorithm and analyze how it affects system performance.

## 2 Related Work

### 2.1 Wearable Sensor-Based Activity Recognition

Much work has been done using BSNs for activity recognition [7], [12]. These BSN-based solutions usually deploy accelerometer sensors on a human body to capture body movements. The sensor nodes are self-organized into a BSN where appropriate MAC and routing protocols are operated to ensure the quality of sensor data. The advantage of BSN-based approaches lies in that they provide fine-grained activity recognition results with high coverage as the user wears the sensor as they move. While BSN-based approaches have shown their effectiveness, some limitations restrict them from being widely adopted for elderly care applications. The users need to correctly wear the sensors and avoiding issues caused by sensor displacement [4]. Due to the sensor nodes are battery powered, constant maintenance to the system such as checking, charging and replacing the batteries is also required which may reduce the practical usability of the system [13].

Different from this approach, we exploit RFID radio information for activity recognition. The advantages of using RFID over BSN sensors are low-cost and easy maintenance [25]. By using low-cost, passive RFID tags instead of active sensor nodes, our system can be embedded into the users' clothes and require no further maintenance. Also, recent study shows it is more energy-efficient to use radio information instead of sensor readings for activity recognition [27].

Recently, smartphone-based activity recognition systems have attracted an increasing research interest [8], [9], [10], [11] to support numerous practical applications [28]. In [8], the authors use smartphone's on-board accelerometer to perform activity recognition. The phone is places in the front pocket of the user's pants during data collection. Zhao et al. [9] propose a cross-people activity recognition system for smartphones with an embedded AI module. Transfer learning technique is used to solve the subject dependency issue of activity recognition systems. In [11], the device orientation and location problem is studied for smartphone-based activity recognition systems. They propose to use signal transformation to achieve accurate recognition with varying phone orientation and location. To sum up, the above work explores the ability of using on-board sensors for activity recognition. To extend the ability of smartphones, some work discusses the combination of wearable technologies with smartphones to construct powerful and efficient activity recognition systems. In [10], the authors use the wearable sensor nodes and the smartphones to achieve practical activity recognition for everyday applications. Their system involves five IRIS sensor nodes to collect movement data from the user's limbs and the head, a smartphone is used to collect sensor data and perform activity recognition. In [29], the authors propose to use MotionBand, a wrist-worn tri-axial accelerometer and a smartphone to collect user's activity data. In [30], Li et al. presents IODetector—a lightweight sensing service on smartphones to detect the indoor/outdoor environments. Their approach uses the smartphone's on-board sensors and assumes no previous knowledge.

Different from the above work, the advantage of our approach is that we use passive RFID technology and radio patterns for activity sensing and recognition which is promising to build a maintenance-free and energy efficient system. As shown later in our experiments, our approach outperforms smartphone-based solution in recognizing fine-grained activities with movements in different body parts.

### 2.2 RF-Based Activity Recognition

Different from wearable sensor-based solutions, RF-based approach has been proposed to perform activity recognition in a device-free manner. In this paradigm, the user is required to wear no or minimum number of devices to perform activity recognition. Instead of using sensor data, fingerprints of RF signal patterns are used to recognize the users' activities.

Some researchers propose to use the wireless signals generated by WiFi APs for activity recognition. In [16], the authors propose CARM for human activity recognition and monitoring using CSI information extracted from two WiFi devices. Base on the sophisticated models that quantify the correlation between CSI dynamics and human activity, their approach achieves real-time and accurate activity recognition in a lab settings. WiSee is presented in [17] to achieve whole-home gesture recognition in a device-free manner. Patterns of Doppler shifts of the wireless signals are captured to recognize the users' activities. Wang et al. presents E-eyes [19] that achieves location-oriented activity recognition in a home settings. Using CSI information extracted from WiFi links, E-eyes conducts activity recognition by comparing channel features with various signal profiles. Adib et al. have conducted extensive research in using WiFi signals to detect the subtle activities of the users involving gesture and moving [31], [32], breathing and heart beating [33]. Detailed information such as the signals strength on different frequency bands are extracted from the reflected WiFi signals from the human body to detect the above subtle activities. In [21], the authors present WiDeo—a device-free motion tracking system which uses standard AP antenna arrays. The amplitude, time of flight, and angle of arrival computed from the reflected signals are used to track the moving objects.

Recently, backscatter technology has attracted much research interest due to its low-cost and low-power properties. Active research has been conducted on high-throughput [34], low-power sensing and communication [35], [36], and large-scale backscatter systems [37]. Much research work has been conducted using backscatter RFID technologies for activity recognition. Early work including LANDMARC [25] and TASA [14] tracks the motion activities of the user by using fixed RFID reader and tag arrays. Their work relies on capturing the patterns of RFID signal RSS affected by the mobility of objects to perform accurate tracking and localization. While early research shows the feasibility of using RFID for detecting gross activities such as motion patterns, recent work explores the ability of this technology to recognize fine-grained user gestures and activities. In [38], the authors use RFID for tracking hand movements in a table-size scale. The tags are placed in a grid-like structure on a table with readers located at three corners. They use tag counting information received at different readers to keep track of hand movements. In [15], the authors propose RF-IDraw that provides accurate tracking of the user's writing gestures in the air using RFID. With a passive RFID tag attached to the user's finger, the system tracks the trajectory of the tag by measuring the signal's angle of arrival with carefully placed antennas of UHF RFID readers. Our previous research [20] also explores to use the variance in RFID signal RSS to recognize user's activities such as falling and posture transitions. Fixed RFID and passive tag arrays are deployed in a home settings to support activity recognition in a device-free manner.

In [27], the authors use the radio communication patterns extracted from a BSN to recognize activities. The communication patterns (i.e., such as packet delivery ratio and the mean of RSSI values) from arrival packets within a time window are extracted and used as a signature to recognize the corresponding activity. In [18], the authors track the motion of objects behind the walls by a self-organized network of radio sensor nodes deployed around the detection area. The RSS of different links across the area of interest is measured and recorded on the base station. The RSS variance caused by moving objects within the covered area is computed and processed using the variance-based radio tomographic imaging algorithm for motion tracking.

The work presented in this paper differs from the above approaches in the following points. First, most existing RF-based approaches assume the availability of fixed radio transmission/receiving devices located near the user [14], [15], [16], [17], [18], [19], [20], [21], [25], [31], [32], [33], [38]. This assumption limits the coverage area of the existing systems considering the mobility of the user. Different from their work, this work achieves high detection coverage by using wearable RFIDs to perform activity recognition. Second, some work relies on specialized devices such as USRP-N210 software radios that capture CSI of wireless links [17], [31] or RFID readers that return the signal phase of each reply [15]. Though such devices are becoming more available than before, the cost and availability of such specialized devices are still prohibitive for everyday usage scenarios. In this work, we choose simple COTS RFID readers that only obtains the RSS information from tag readings to provide a lower-cost and more accessible solution. Finally, different from the wearable solution that relies on BSN nodes [27], this work explores the use of passive RFID technology for activity recognition which requires less maintenance effort.

### 2.3 Real-Time Activity Recognition

Many previous systems perform activity recognition in an off-line manner [7]. Off-line recognition algorithms cannot achieve real-time recognition because they require the complete activity trace to perform recognition. The data collection time of such off-line recognition systems, which is also a part of the system's recognition latency, is long and uncontrolled. As a result, a real-time activity recognition system must be able to perform on-line activity recognition [39]. Different from off-line approaches, our system is can perform accurate activity recognition in real-time. Real-time activity recognition systems are more practical in supporting everyday applications such as emergency response and smart reminder.

In [39], the authors discuss the performance of a real-time activity recognition system that considers both the recognition accuracy and recognition latency given two delay bounds (without and with quality degrade). They propose a hierarchical, online activity recognition system that can perform best-effort soft real-time activity recognition. They system achieves an average recognition delay of 5.7 sensing periods and an average recognition accuracy of 82.87 percent. Different from this work, we discuss the problem of real-time activity recognition with a single delay bound which must be achieved.

## 3 Preliminary Experimental Studies

In this section, we introduce our system's hardware setup and conduct preliminary experiments to show the tag reading performance under different conditions and the potential of using radio patterns for activity recognition.

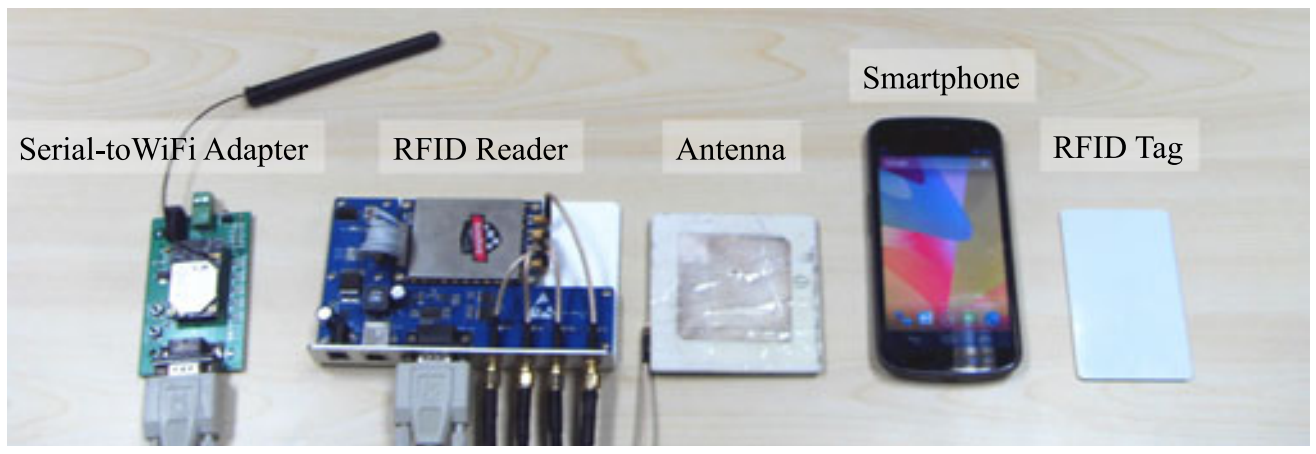

Fig. 1. Hardware setup.

### 3.1 Hardware Setup

The hardware used in this work is shown in Fig. 1. We use an Impinj R2000 RFID reader module[1] powered by a Li battery with 6800 mAh capacity (matchbox size, 3 cm × 6 cm × 2 cm). The size of the reader is 15 cm × 9 cm × 2.5 cm. We use UHF RFID tags with a credit card size. We have four antennas, and each has a size of 7.8 cm × 7.8 cm × 0.5 cm. The transmission power level of each antenna is adjustable from 0 to 30 dbm with a minimum level of 0.1 dbm. The RFID reader module operates at 840–960 MHz and supports UHF RFID standards including US FCC 47 CFG Ch. 1 Part 15 and ETSI EN 302 208-1. The agility of the module is −95 dbm. When set to the tag inventory mode, the reader can read as many tags as possible (maybe multiple readings per tag) using an anti-collision protocol. For the reader we used, over 50 tag readings can be obtained in one second. Each tag reading contains the tag ID[2] and the RSS value. Tag readings obtained from the reader are sent wirelessly through a Serial-to-WiFi adapter to a smartphone for further processing. We use Samsung Nexus 3 smartphone[3] with a due-core 2.4 G processor, running Android 4.0.

### 3.2 Reading RFID Tags

We first study the tag reading performance of the reader under different settings in transmission power level, tag-antenna distance, orientation, and the presence of human body.

First, Fig. 2a shows the RSS values obtained at different positions in the detection area of an antenna. The antenna is placed on the top of the area facing downward with the transmission power level set to 20 dbm. The tag under test is placed in an area 0 to 240 cm perpendicular to, and −120 to 120 cm parallel to the antenna face (we use negative values to represent positions on the right side of the antenna). The tag-antenna angle is 0°, i.e., the tag face is parallel to the antenna face. As we can see from Fig. 2a, the RSS gets stronger when the tag is placed closer to the antenna. Specifically, in the direction perpendicular to the antenna face, the tag can be stably read when placed within the distance of 60 to 90 cm. For a distance less than 60 cm or larger than 90 cm, the tag is not detected in some locations. In the direction parallel to the antenna face, the tag can be stably read when placed in the distance of −60 to 60 cm. For locations out of this range, the tag is not detected sometimes.

Next, we change the antenna's transmission power level to 30 dbm and repeat the previous experiment. The results

1. http://www.impinj.com/products/reader-chips/
2. We use the Electronic Product Code (EPC) stored on a tag as its ID.
3. http://www.samsung.com/us/mobile/cell-phones/SPH-L700ZKASPR

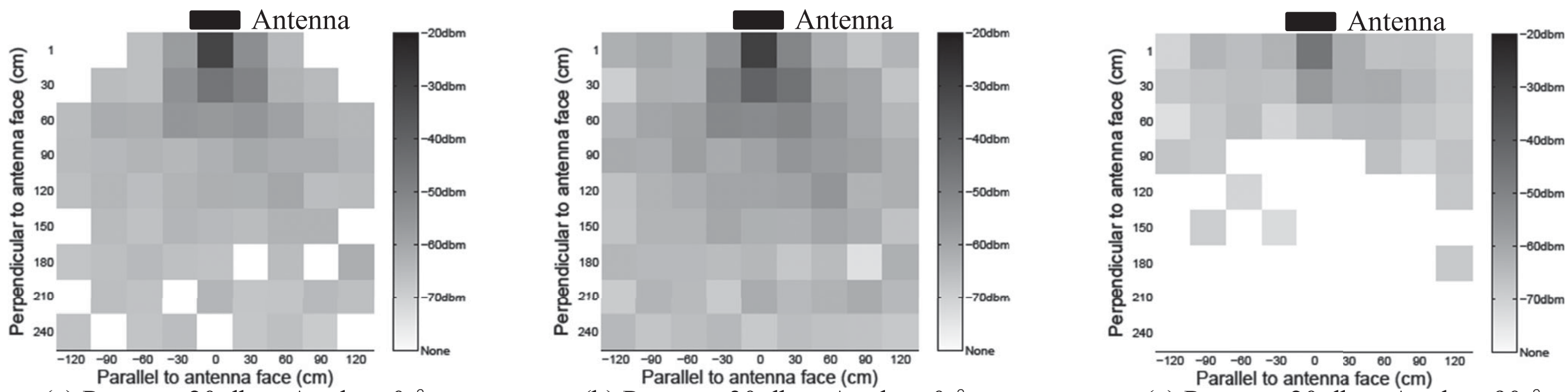

(a) Power = 20 dbm, Angle = 0 ° (b) Power = 30 dbm, Angle = 0 ° (c) Power = 20 dbm, Angle = 90 °

Fig. 2. Average RSS of tags at different positions with different transmission power levels and tag-antenna orientations.

are shown in Fig. 2b. It is clear that all the RSS values get increased as compared to the 20 dbm results in Fig. 2a, and the antenna's reading range covers the entire 240 cm × 240 cm area with no miss detection.

Again, we repeat the first experiment with a transmission power level of 20 dbm, but we turn the tag-antenna orientation from 0 to 90°. The results are shown in Fig. 2c. From the figure, we observe the antenna's reading area has changed significantly. Tags can be read farther in the positions parallel to the antenna's face compared to Fig. 2a. However, in the direction perpendicular to the antenna's face, the reading distance is significantly reduced to no more than 60 cm. While the rule of closer tags get higher RSS values still holds, by comparing Fig. 2a and Fig. 2c, we can see that the RSS values are lower after turning the tag 90°, especially for positions directly in front of the antenna (i.e., 0cm distance parallel to the antenna).

Finally, since the human bodies are known to absorb RFID signals [40], we study the influence of different body parts to tag readings by blocking the line-of-sight between the tag and the antenna. The results are shown in Fig. 3. The antenna and the tag are placed fact-to-face with a fixed distance of 0.5 m. As we can see from the figure, the body completely blocks the signal when transmission power level is lower than 20 dbm, the arm and the leg also blocks the signal when transmission power level is lower than 12.5 and 15 dbm, respectively. For transmission power level higher than 20 dbm, the signal is not completely blocked but significantly weakened. The body, which is the largest part, reduces the RSS values by approximately 20 dbm. The arm and the leg, which are similar in sizes, reduces the RSS values by approximately 10 dbm with the leg slightly higher than the arm.

In summary, considering the height of a human is normally within 2 m, we conclude that the reading range of our device can cover the whole human body with a proper transmission power level such as 20 dbm. With a certain transmission power level, the tag RSS values are affected by factors including tag-antenna distance, orientation, and body blockage. By wearing the tags and antennas on the user's body, the above factors will change by the movements of different body parts when performing activities, and can be used for activity recognition.

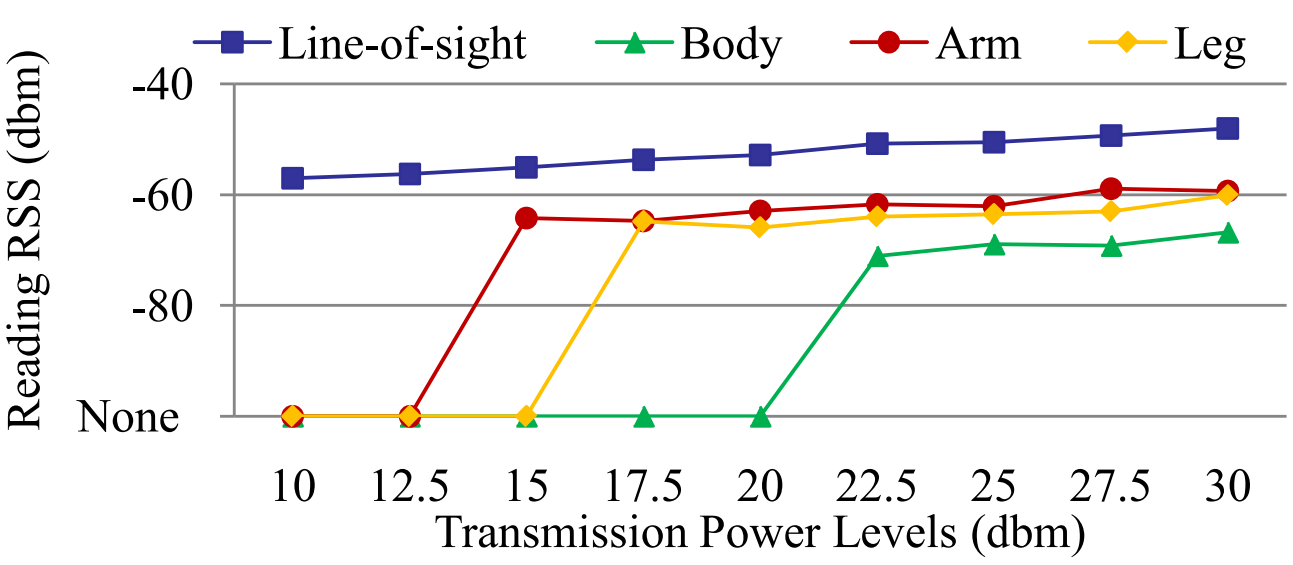

Fig. 3. Influence of human body to tag readings.

### 3.3 Potential of Activity Recognition

In this section, we demonstrate the potential of using RFID radio patterns extracted from RSS values for activity recognition. The experiment is carried out by one male subject performing three basic motions including standing, sitting, and walking. Four antennas are attached to the body located at the chest, back, left foot, and the right foot. Tags are attached to nine body parts including the left/right wrists, left/right arms, body, left/right legs and left/right ankles to capture body movements. The details of such tag and antenna placement strategy will be introduced later in Section 4.1. During this experiment, we use the all antenna inventory mode of the reader which automatically activates the antennas for tag reading. Under this mode, the readings from the four antennas are mixed together. We attach four tags around each body part to increase reliability. The transmission power level of the RFID reader is set to 30 dbm. In this experiment, the four tags attached around the same body part share the same tag ID. We use the above setting in this experiment to demonstrate the discriminative power of RSS values for activity recognition.

The average RSS values of tags attached to different body parts are shown in Fig. 4. As shown in this figure, it is clear that different motions result in different RSS patterns of tags. For the sitting activity, the RSS values of tags attached to the body, right leg, and left/right ankles are clearly stronger and more stable than other activities. It is possibly because when sitting, the user is stationary and the tags attached to the above body parts are closer to the antennas than other activities. For the standing activity, the RSS readings of the above mentioned body parts are also relatively more stable than the walking activity but the RSS values are lower than the sitting activity. Additionally, it can be seen from Fig. 4 that the RSS values of tags attached to the left wrist and left arm are stronger and more stable than other activities. For the walking activity, rhythmic variances in RSS values can be observed for tags attached to nearly every part of the human body, and they seem matched with arms and legs waving during walking.

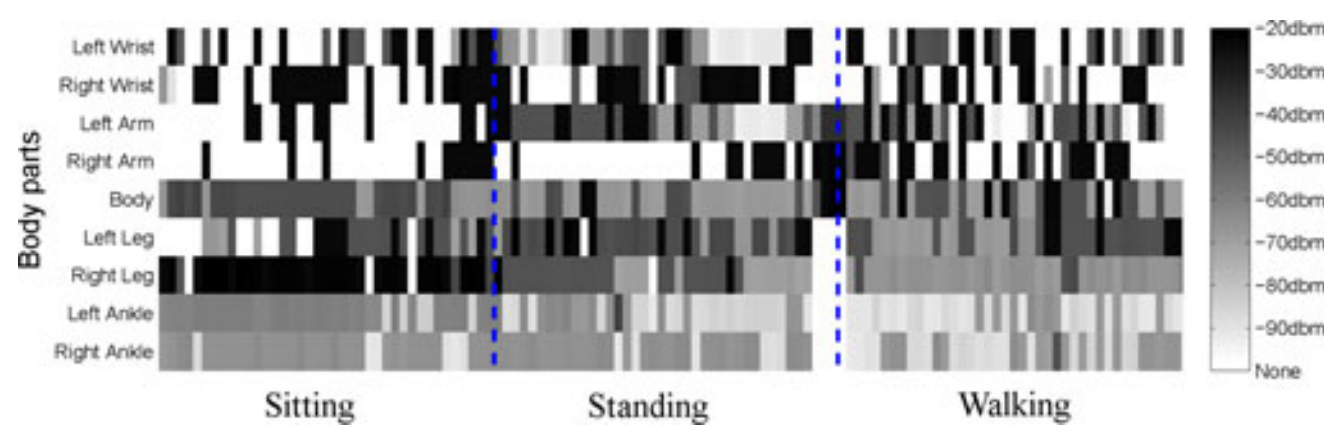

Fig. 4. Average RSS with different activities over time.

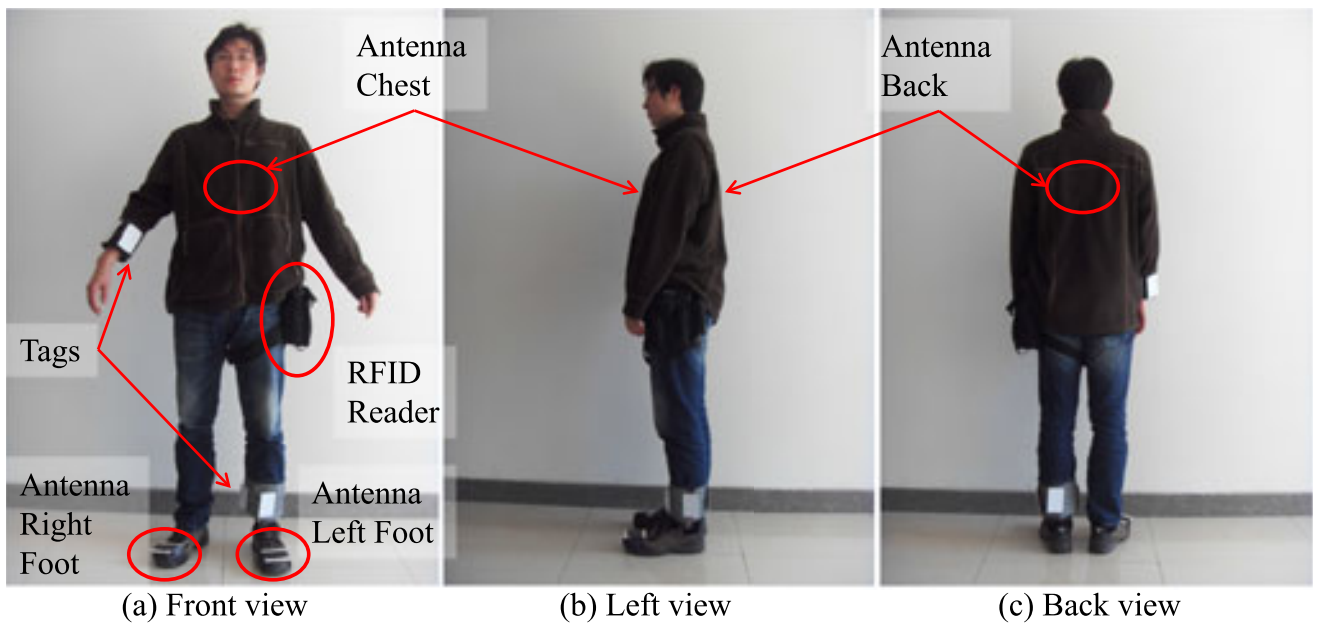

Fig. 5. Subject wearing the RFID reader and antennas, part of the tags are shown while others are hidden in the clothes.

Our preliminary studies have shown the feasibility and potential of using RFID radio patterns for activity recognition. In the next section, we present our system design in detail.

## 4 System Design

In this section, we present the detailed design of the proposed RFID-based real-time activity recognition system.

### 4.1 Antenna / Tag Placement

We present the antenna and tag placement strategies in this section and show a subject wearing our prototype in Fig. 5.

*Antenna Placement*. As suggested by Fig. 2 in our preliminary studies, we place four antennas on a human body – two antennas (one on the chest and the other on the back) for detecting hand/arm movements, and one antenna on each of the feet for detecting lower body movements (as shown in Fig. 5). Such placement ensures a total coverage of different body parts, and also meet user's comfort need.

*Tag Placement*. To capture the movement of different body parts, RFID tags are attached to nine body parts including both wrists, arms, ankles, legs, and the body. To increase the reliability of tag readings, we attach four tags at each body part. For example, for the right wrist, we attach four tag located at the front, left, right, and back of the wrist. This redundant tag placement strategy ensures that no matter how the user moves his/her wrist, at least one tag will face the antenna and can be read by the reader with high probability. A total number of 36 tags are attached on the user's body, with each tag having an unique ID.

*Inventory Mode*. Instead of using the all antenna inventory mode used for our preliminary experiment, we use single antenna inventory mode to discriminate the readings of one antenna from others. The four antennas connected to the RFID reader are activated sequentially to detect tags within their reading ranges. The dwell time of each antenna is set to two seconds and the time to complete an inventory cycle is eight seconds. We choose this value because it achieves a balance between reading stability and agility. The tag readings obtained during the activation time of an antenna is a series of tag IDs and their RSS values in the following format

$$< timestamp, antennaID, tagID, RSS > .$$

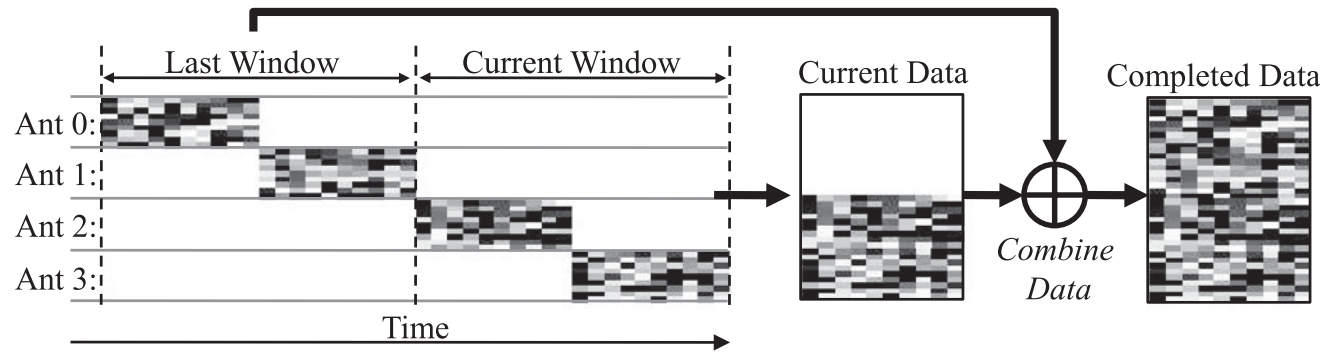

Fig. 6. Data completion method.

The transmission power level of the RFID reader is a key parameter in our system, and it influences the system's performance on both recognition accuracy and battery consumption. We find the optimal transmission power level by experiments in Section 5.4.2. We also evaluate the effect of different antenna and tag placement strategies in Section 5.4.1.

### 4.2 Data Segmentation and Completion

Given the continuous stream of tag readings, we first apply a sliding window to segment the data. The window length is specified by the parameter $L$ in seconds. $L$ is a key parameter in our system for it affects both the recognition accuracy and latency. In [27], the authors propose an adaptive window size adjustment algorithm to find the optimal sliding window size. In this paper, however, we focus on real-time activity recognition with a restricted recognition latency defined by the application. As a result, we use a fixed window size specified by $L$ combined with the real-time activity recognition algorithm introduced later to achieve stable time performance. We evaluate the system's performance under different sliding window sizes later in Section 5.

As mentioned in the introduction, one of the challenges in existing RFID systems is false negative readings [14], [38], [41], caused by miss detection – a tag is in the antenna's reading range, but not detected. In addition, in our system if the sliding window size is too short for the reader to complete readings for all four antennas, it may also cause false negative readings. To address this issue, we use recent historical data to complete the current readings. We illustrate the method in Fig. 6, assuming the false negative reading is caused by a short window size of 4 s. While the data from antenna 0 and 1 are missing in the current data because the current window is only long enough to complete two antennas' readings, we use the last window's data to complete the current data. The intuition behind this approach is *temporal locality* – tags recently detected are likely to be detected again with similar RSS values. While the idea is simple, it is important to determine how much historical data to use when trying to complete the current data. The history should be long enough to complete the missing part but not too long to overwhelm the current data.

We keep a runtime buffer that stores the data collected in at most the last 20 s (long enough for the reader to complete two cycles of inventory on all the antennas and not too long considering the average duration of an activity [7]) as history. More specifically, for a window size of $L$ seconds and the current data segment obtained at time $t$, $D(t)$, the buffer contains a series of recent data segments $<D(t-\lfloor\frac{20}{L}\rfloor L)$, $D(t-\lfloor\frac{20}{L}\rfloor L+L),\ldots,D(t-L)>$. Given the current data $D(t)$, we start data completion by looking into the data obtained most recently, $D(t-L)$. We keep monitoring tag count distributions of the historical data obtained at time $t'$, $D(t')$, and the current data (maybe partially completed already), $D(t)$, and terminate the completion process as soon as they overlap too much to prevent overrun. The procedure is sketched in Algorithm 1 where the *overlap* function used in line 3 quantifies the overlap in tag count distributions of current data $D(t)$ and historical data $D(t')$ as follows

$$overlap(D(t),D(t'))=\frac{\sum_{i=1}^{N_a}\sum_{j=1}^{N_t} min(count_{i,j}(t),count_{i,j}(t'))}{\sum_{i=1}^{N_a}\sum_{j=1}^{N_t} count_{i,j}(t')}.$$

where $N_a$ and $N_b$ are the total number of antennas and tags in the system, respectively, and $count_{i,j}(t)$ is the count of tag $j$ read by antenna $i$ at time $t$. In line 3, an empirical overlap threshold of 0.7 is used to terminate the completion process.

**Algorithm 1.** Data completion.

**Input:** Current data $D(t)$;
History data $<D(t-\lfloor\frac{20}{L}\rfloor L),D(t-\lfloor\frac{20}{L}\rfloor L+L),\ldots,D(t-L)>$.
**Output:** Current data $D(t)$ with missing readings filled.

1: $n=1$
2: **while** $t-nL\geqslant t-\lfloor\frac{20}{L}\rfloor L$
3: **if** overlap($D(t)$, $D(t-nL)$) $<0.7$
4: Append $D(t-nL)$'s readings to the front part of $D(t)$
5: $n=n+1$
6: **else break**
7: **end-if**
8: **end-while**
9: **return** $D(t)$

The above method is highly effective in improving recognition accuracy as we will show later in our experiments (Section 5.6). The experimental results also suggest that the overlap threshold of 0.7 we choose is optimal on our data set.

## 4.3 Temporal and Spatial Features

For each data segment, we extract both temporal and spatial features to characterize the radio patterns. However, we also need to limit the dimension of the feature set since we target for a real-time system. Moreover, readings from different combinations of tags and antennas may be different even with the same condition. This is an known performance issue of tag readings commonly exists in RFID-based systems [14], [25]. This issue may significantly affect the quality of the features if not carefully treated.

### *4.3.1 Temporal Features*

The data in each segment are composed of series of RSS values arranged in an increasing order with each series representing the RSS values of a specific tag read by a specific antenna. Seven features are extracted from each RSS series including the mean, variance, max, min, mean crossing rate, frequency domain energy and entropy of the RSS values to characterize its radio patterns temporally. The mean, max, min, mean crossing rate, and variance of RSS values are used to characterize the overall strength and stability of the signal. The frequency domain energy captures the periodicity of the data and entropy is used to help discriminate activities with similar energy values [7]. Different from work [14], [25] that use readings from different tags and antennas cooperatively, our temporal features are extracted for each RSS series independently from the others. The performance issue of tag readings may cause tags attached to similar body parts to have different RSS values but will not affect the discriminative power for each RSS series has its own different radio patterns for different activities.

The 4 antennas and 36 tags attached to the user's body generates a total number of 144 RSS series in a data segment. As a result, there are 1008 temporal features extracted from each data segment.

### *4.3.2 Spatial Features*

To characterize the radio patterns spatially, we extract the correlation coefficients of RSS series for different tags read by different antennas. The correlation coefficient quantifies the degree of dependency between a pair of RSS reading series by observing the similarity in their changing patterns. As shown by our preliminary experiment results, the RSS values are stronger with a closer tag-antenna distance and a smaller tag-antenna angle when the transmission power level is fixed. The blockage of human body also have influence to RSS values. The performance issue of tag readings may cause different RSS values obtained by different combinations of tags and antennas even with the same condition but cannot fundamentally change their changing patterns. For our solution, we leave out features that are sensitive to the performance issue such as mean and variance across different RSS series to avoid calibration as did in [14], [25].

Ideally, the correlation of each pair of RSS series should be computed, leading to a total number of 10,296 correlation coefficients. However, it may incur a large computational overhead. As a result, we trade feature quality with dimensionality by computing the correlations between tags and antennas separately. For each tag, we select a representative reading series by choosing the one with the largest reading count across all four antennas. We then compute the correlation between each pair of tags using their representative readings. These correlations roughly characterize the dependency relationship between RSS readings of different tags when assuming the antennas are fixed in positions. Similarly for each antenna, we select the readings series with the largest reading count across all 36 tags as the representative readings and compute the correlation between each pair of antennas. These correlations then roughly characterize the dependency relationship between RSS readings obtained by different antennas when assuming the tags are fixed in positions.

By using the above technique, the spatial features are reduced to 636. After feature extraction, an activity instance is obtained and ready for recognition.

## 4.4 Real-Time Recognition Algorithm

The design goal of our system is a real-time activity recognition system, which completes recognition within a delay bound. We identify two key requirements described as follows.

TABLE 1
Four Subjects Involved

| No. | Gender | Height | Weight | Shape |
|---|---|---|---|---|
| 1 | Male | 177 cm | 70 kg | Normal |
| 2 | Female | 159 cm | 45 kg | Slim |
| 3 | Male | 180 cm | 75 kg | Normal |
| 4 | Male | 193 cm | 110 kg | Strong |

TABLE 2
Eight Activities Studied

| No. | Activity | No. | Activity |
|---|---|---|---|
| 1 | Sitting | 5 | Cleaning Table |
| 2 | Standing | 6 | Vacuuming |
| 3 | Walking | 7 | Riding Bike |
| 4 | Cleaning Window | 8 | Going Up/Down Stairs |

*Online*. An recognition algorithm is offline if it requires the complete instance of an activity to be presented for recognition [7]. Offline systems cannot perform real-time recognition for they need to wait for the current activity to finish before recognition and the waiting time is uncertain. To achieve real-time recognition, the algorithm must be online that can recognize the current activity without the complete activity instance, i.e., only using data already obtained.

*Continuous*. To achieve real-time, the recognition result must be obtained before the delay bound. Since the recognition system works iteratively to generate recognition results, we adapt the real-time concept from the signal processing field [42] to activity recognition systems. The acceptable recognition latency is specified by the sliding window size $L$ which determines the data collection time. The processing time must be less than the data collection time [42] so that the recognition result can be obtained before the next data segment arrives, providing continuous recognition results without extra delays.

While the online property of our recognition system is guaranteed in our system for activity instances are generated only using the data already obtained, the continuous property is determined by the execution time of the recognition algorithm. We design a fast recognition algorithm based on a multi-class support vector machine (SVM) with radial basis function kernel. SVM is widely used in activity recognition [27], [43]. The advantage of using SVM for activity recognition includes: 1) designed on a sound theoretical basis, SVM is promising to have accurate and robust classification results; 2) SVM scales well to the number of features; 3) the model training can be performed on very few training cases; and 4) the recognition can be executed fast at runtime [27]. We have implemented the recognition algorithm on a smartphone, and evaluate its real-time performance in the next section.

### 4.5 Antenna and Tag Selection

As introduced previously in Section 4.1, a number of antennas and tags are placed on the user's body to capture radio patterns for different activities. This strategy enables collection of extensive RSS readings which capture the movements of different body parts, however, users may feel uncomfortable if wearing too many such devices, especially antennas which are relatively larger in size and heavier than tags. Aiming to minimize the number of antennas and tags, in this section, we present a wrapper algorithm for antenna and tag selection that searches for the minimum number of antennas and tags required to achieve an acceptable recognition accuracy.

Given an accuracy threshold $\rho$, the algorithm starts with one antenna and one body part (involves four tags) and tests the proposed SVM-based recognition algorithm. If the recognition accuracy $a < \rho$, then the algorithm increases the number of antennas and body parts selected until the condition $a \geqslant \rho$ is met. The complete algorithm is shown in Algorithm 2.

In Algorithm 2, two variables $n_{ant}$ and $n_{tag}$ indicate the number of antennas and tags selected, and the operations $\binom{Ant}{n_{ant}}$ and $\binom{Ant}{n_{ant}}$ choose $n_{ant}$ antennas from the antenna set $Ant$ and $n_{tag}$ tags from the tag set $Tag$, respectively. The algorithm may output more than one result if the accuracy threshold is achieved by different sets of antennas and tags with the same antenna and tag number.

**Algorithm 2.** The Antenna and Tag Selection Algorithm.

**Input:** Accuracy threshold $\rho$, the set of antennas $Ant$, the set of tags $Tag$.
**Output:** The minimal set of antennas and tags required to meet the accuracy threshold.

```
1:  flag = false, n_ant = 1, n_tag = 1
2:  while flag == false && n_ant ⩽ |Ant|
3:    while flag == false && n_tag ⩽ |Tag|
4:      for each Ant_s ∈ (Ant choose n_ant)
5:        for each (Tag_s ∈ (Tag choose n_tag))
6:          if Accuracy(Ant_s, Tag_s) = a ⩾ ρ then
7:            flag = true
8:            output Ant_s, Tag_s, and a
9:          end-if
10:       end-for
11:     end-for
12:     n_tag = n_tag + 1
13:   end-while
14:   n_ant = n_ant + 1, n_tag = 1
15: end-while
```

## 5 Empirical Studies

In this section, we present empirical studies to evaluate the performance of our system.

### 5.1 Data Collection

The experiments are conducted in an area of our office building, including two rooms and a corridor, as well as outdoors. Our data collection involves four subjects (three males and one female). The subjects are carefully selected to represent different heights and body types as summarized in Table 1. We use three transmission power levels (i.e., 20, 25, and 30 dbm) according to the result from our preliminary studies. Each subject is required to perform eight activities as summarized in Table 2, and a snapshot of these activities is shown in Fig. 7. The data collection is carried out over a period of two weeks and a total number of over 200 activity instances are collected.

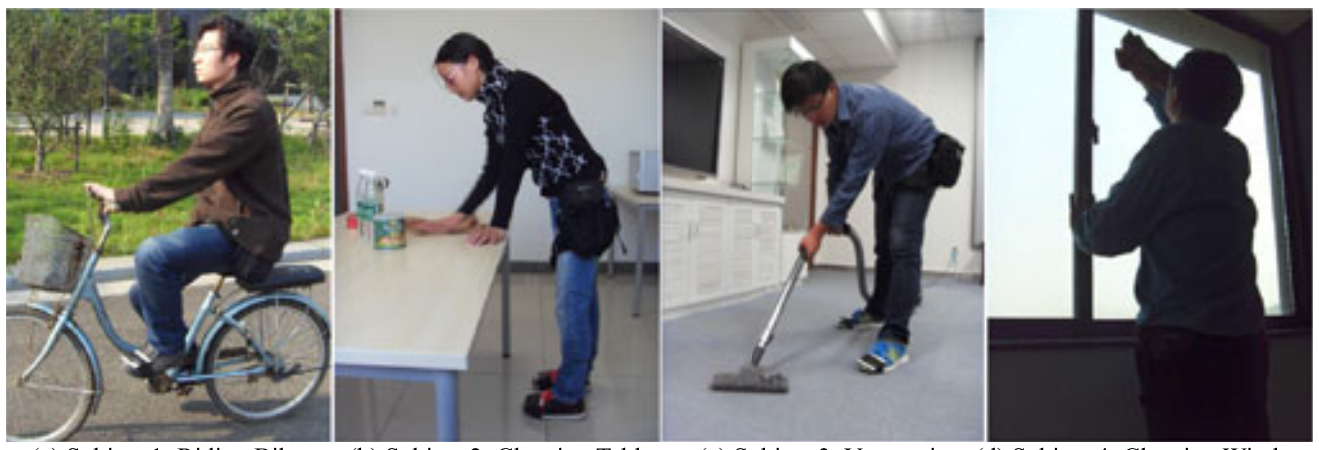

(a) Subject 1: Riding Bike (b) Subject 2: Cleaning Table (c) Subject 3: Vacuuming (d) Subject 4: Cleaning Window

Fig. 7. Data collection.

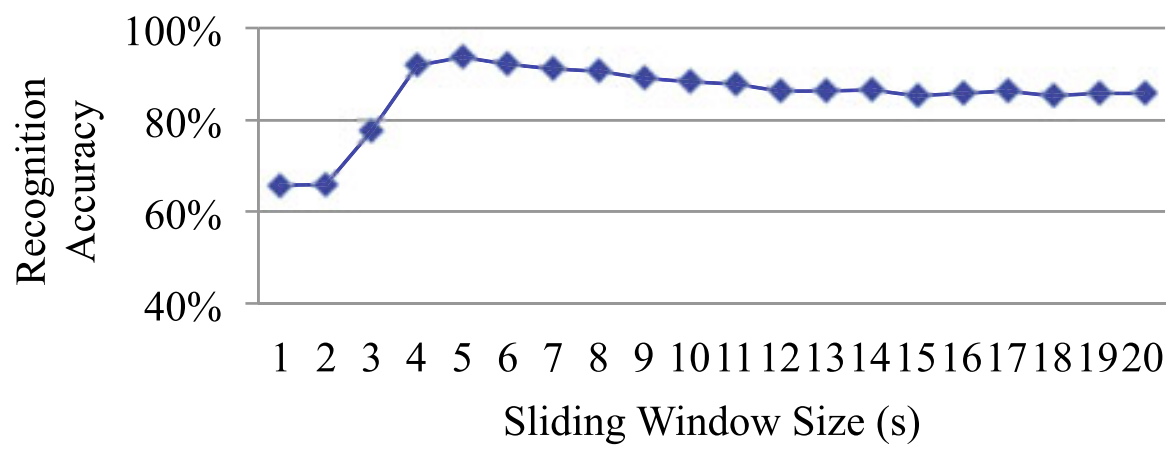

Fig. 8. Sliding window size versus overall recognition accuracy.

## 5.2 Recognition Performance

### 5.2.1 Recognition Accuracy and Latency

In the first experiment, we use ten-fold cross-validation to evaluate the overall recognition accuracy which is defined by the number of correctly classified instances over the number of the total instances, and the latency which is determined by the sliding window size. The definition of overall recognition accuracy is as follows

$$overall \quad recog. \quad accracy = \frac{\#Correct \quad Inst.}{\#Total \quad Inst.}.$$

The recognition accuracies with different sliding window sizes are illustrated in Fig. 8. As we can see from the figure, when the sliding window size is small (i.e., from 1 s to 4 s), the recognition accuracy rapidly grows from 65.8 to 91.8 percent. The recognition accuracy reaches its peak at 93.6 percent when sliding window size is 5 s. It is interesting to see that the recognition accuracy drops slowly afterwards and stabilizes at around 86 percent when the sliding window size increases further. This result suggests that a larger sliding window does not always result in a higher recognition accuracy. This may be due to a larger sliding window size increases the chance of getting different activities' data into one sliding window.

We breakdown the recognition performance of different activities with a window size of 5 s and show the results in Fig. 9. We use precision and recall defined as follows as metrics[4]

$$precision = \frac{TP}{TP + FP} \quad recall = \frac{TP}{TP + FN}.$$

Fig. 9 shows that the precision and recall for most of the activities are above 0.9. By analyzing the results, we find that some of the *walking* activity are recognized as *going up/down stairs*, a few instances of the *walking*, *cleaning table*, and *cleaning window* activities are recognized as *vacuuming*.

4. We use TP, FP, TN, FN for true positive, false positive, true negative, and false negative, respectively.

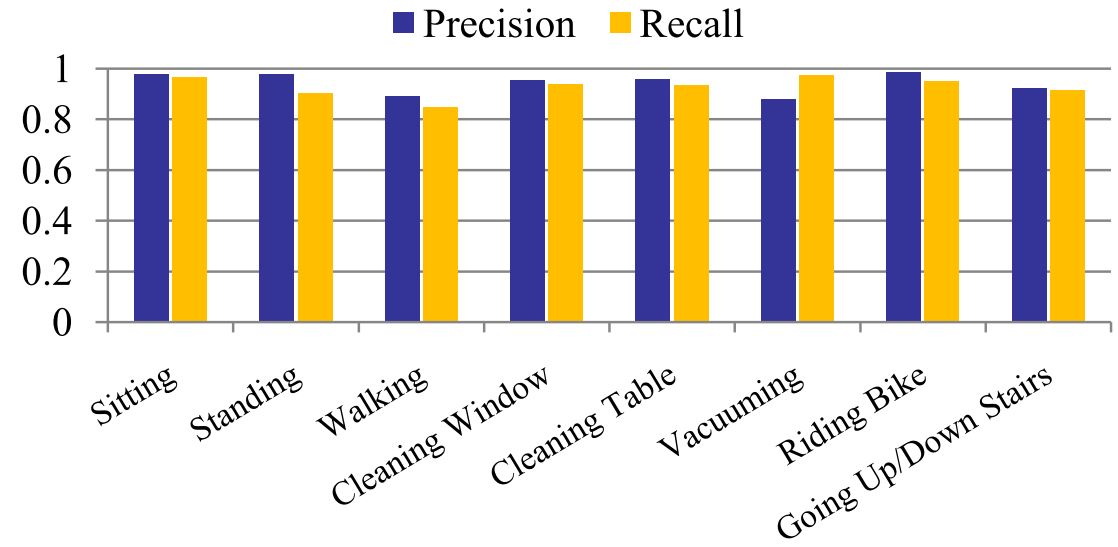

Fig. 9. Breakdown of Precision and recall.

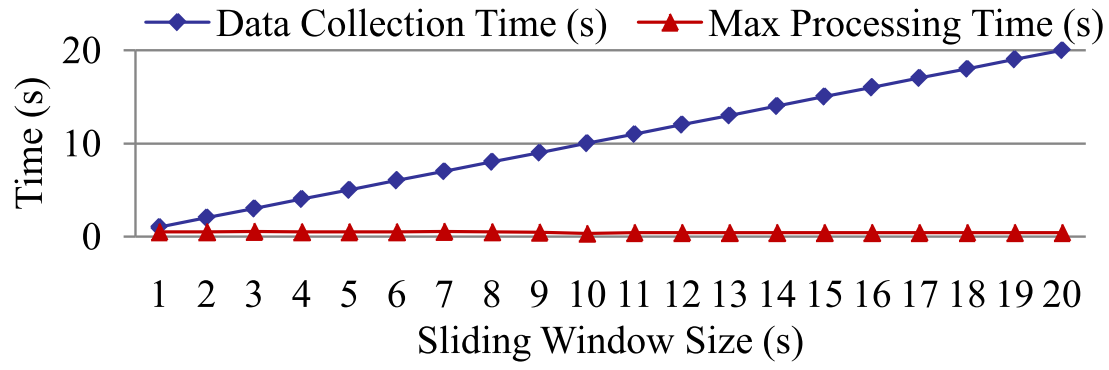

Fig. 10. Data collection time versus maximum processing time.

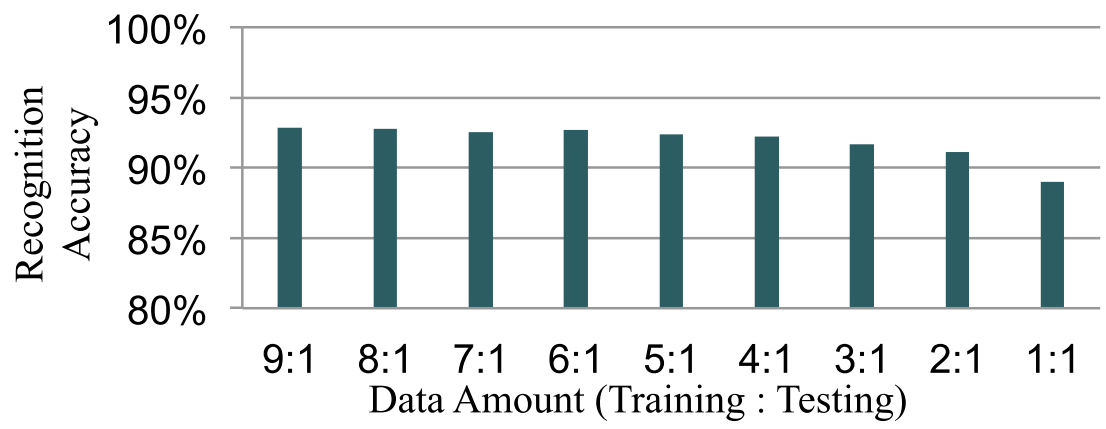

Fig. 11. Accuracy with different training data amount.

Overall, as shown in Fig. 8, our system achieves the best overall recognition accuracy of 93.6 percent when the sliding window size is set to 5 s. We use a sliding window size of 5 s for the rest of our experiments.

### 5.2.2 Real-Time Performance

In this section, we evaluate the real-time performance of the system. The online property is guaranteed by using only the current and historical data for recognition. The continuous property is determined by the execution time of the recognition algorithm. Fig. 10 compares the data collection time and the maximum processing time on our smartphone (data completion + feature extraction + recognition) under different sliding window sizes.

As shown in the figure, our system performs real-time recognition even when the delay bound is down to 1 s (by fixing the sliding window size to 1 s). The maximum processing time is always less than the data collection time and remains low (around 450 ms) when the sliding window size grows.

## 5.3 Training Data Quality

In this experiment, we evaluate how the quality of training data affects the recognition results.

### 5.3.1 Amount of Training Data

We first study how the amount of training data affects recognition accuracy. Fig. 11 shows the recognition accuracy when taking different portion of the data set as training data. The recognition accuracy is 89 percent when the ratio between data for training and testing is 1:1 (50 percent data for training and the rest for testing). The accuracy increases rapidly with the increase of training data amount and rises above 92 percent after the ratio of training data is above 4:1. This result suggests that the proposed system can achieve good performance without a large amount of training data.

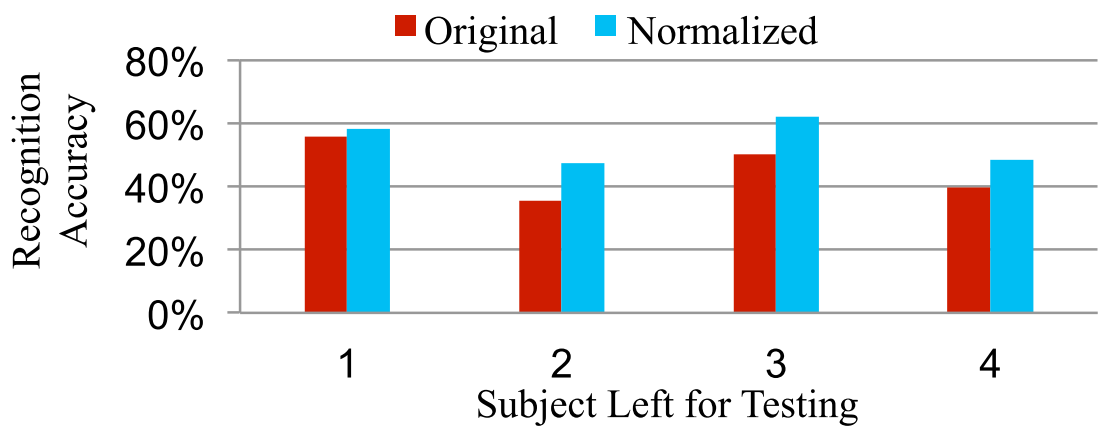

Fig. 12. Subject dependency, the red and blue bars show the recognition accuracy when using the original and normalized RSS values, respectively.

#### 5.3.2 Subject Dependency

Besides training data amount, another important factor for training data quality is subject dependency. We use leave-one-subject-out evaluation to study the our system's performance. Fig. 12 shows our results. The red bars show the recognition accuracy when using the original RSS values for feature extraction. This result shows the recognition accuracy is reduced to an average of 45 percent, suggesting a general dependency on subjects. The recognition accuracy for subject 2 is the lowest. Referring to the subjects' information shown in Table 1, a possible explanation for this result is that different subjects absorb different amount of signal energy because of differences in body mass and height (further supported by optimal transmission power for different subjects in Section 5.4.2).

One possible way to eliminate such differences is to normalize the RSS values before feature extraction. The blue bars in Fig. 12 show the corresponding recognition accuracy. An average of 9 percent increase is observed. For subjects 2 and 3, we observe an increase of 12 percent in recognition accuracy.

Though the experiment result suggests our system is subject dependent, we demonstrate that the performance can be significantly improved by a simple technique. Moreover, since our system can achieve good performance without a large amount of training data, the proposed technology is still promising for real-world applications.

#### 5.3.3 Location and Time Dependency

To evaluate the location and time dependencies of our approach, we test the system's performance using training and testing data collected from different locations and time. For location dependency, we collect data in two different places. The training data are collected in a lab room which is nearly empty only with two tables and a few chairs necessary to perform the activities. The testing data, on the other hand, are collected in a crowded office room with multiple students working in it and many reflection surfaces. The testing result shows that the recognition accuracy of our system achieves 82.8 percent. For time dependency, we train our model with the training data and test the system's performance with testing data collected 1 day, 4 days, and 12 days later. The recognition accuracies are 82.1, 76.2, and 76.1 percent, respectively.

TABLE 3
Antenna Placement Configurations

| Configuration | Back | Chest | Left Foot | Right Foot |
|---|---|---|---|---|
| Upper Antennas | ✓ | ✓ | | |
| Lower Antennas | | | ✓ | ✓ |
| Mixed Antennas | | ✓ | ✓ | |

From the above results, we observe that variation exists in recognition accuracy for both location and time dependency. This may be due to radio signal propagation behavior is much dependent on the surroundings and environment, and hence can be highly variable. In our future work, we plan to investigate this issue further and use transfer learning [44] to address the variation due to location and time dependence.

### 5.4 System Design Choices

In this experiment we evaluate how different system design choices affect the system's performance.

#### 5.4.1 Antenna and Tag Placement

In this section, we first present a manual selection approach to test the recognition accuracy under different sets of antennas and tags. We then obtain more comprehensive results by the proposed antenna and tag selection algorithm which requires long execution time due to the large search space.

*Manual Selection*. In this experiment, We manually select the antenna and tag placement strategies and evaluate their performance. We first divide the antennas and tags into different groups and test the system's recognition accuracy under different configurations. We designed three placement configurations for both the antennas and the tags as shown in Tables 3 and 4, respectively. Note that we assume the user wears all the tags when choosing different antenna configurations and wears all the antennas when choosing different tag configurations. The heuristic behind these strategies is: the antennas and tags placed on the upper (lower) body are more sensitive to the RSS variances caused by upper (lower) body movements.

The recognition accuracies under different antenna configurations are illustrated in Fig. 13. To compute the accuracy of each activity, we use the same metric defined in [27]:

$$per\quad activity\quad accuracy = \frac{TP + TN}{TP + TN + FP + FN}.$$

TABLE 4
Tag Placement Configurations

| Configuration | Left Wrist | Right Wrist | Left Arm | Right Arm | Body | Left Leg | Right Leg | Left Ankle | Right Ankle |
|---|---|---|---|---|---|---|---|---|---|
| Upper Tags | ✓ | ✓ | ✓ | ✓ | ✓ | | | | |
| Lower Tags | | | | | ✓ | ✓ | ✓ | ✓ | ✓ |
| Mixed Tags | | ✓ | | ✓ | ✓ | ✓ | | ✓ | |

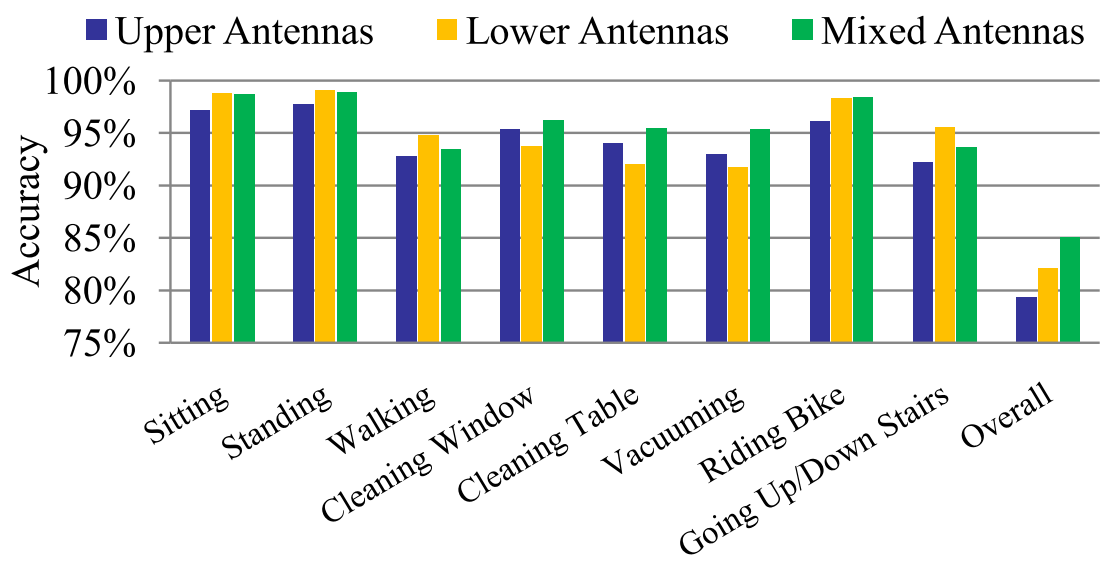

Fig. 13. Accuracy under different antenna configurations.

The metric for the overall accuracy is the same as we used in our first experiment. Fig. 13 shows that the lower antennas are effective to activities involving more lower body movements (e.g., *sitting*, *standing*, and *walking*), and the upper antennas are more effective to activities (e.g., *cleaning window*, *cleaning table*, and *vacuuming*) with more upper body movements. The *Mixed Antennas* configuration achieves the highest overall accuracy of 85.1 percent.

For different tag configurations, the results are illustrated in Fig. 14. We have similar observations as in the antenna configuration experiment. The *Lower Tag* and the *Mixed Tag* configurations achieve similar overall accuracy of 88.8 and 87.8 percent, respectively.

*Search Results*. In this experiment, we present the search results of the minimum number of antennas and tags required to achieve a pre-defined accuracy threshold $\rho$ using the antenna and tag selection algorithm (Algorithm 2). Though Algorithm 2 can be applied to perform fine-grained antenna and tag selection, in this experiment, we choose the set of tagged body parts to reduce the time cost.

The minimum number of antennas and tagged body parts required under different $\rho$ values are shown in Fig. 15. This result suggests that two antennas are sufficient to achieve high recognition accuracy. More tags are required to be placed on the user's body to achieve higher accuracy. From Fig. 15 we discover that the number of antennas and tags required are much fewer than we initially placed on the user's body. For example, to achieve a reasonably high recognition accuracy of 85 percent, we only need two antennas and four tagged body parts.

To achieve the minimum number of antennas and tags, careful selection is required. For example, Table 5 shows the four cases of minimum number of antennas and tagged body parts required when $\rho = 85\%$. The results suggest it is good practice to choose one antenna from the upper body and the other from the lower body. The antennas placed on the user's chest and the left foot are often more useful than antennas placed on the back and the right foot. For tags, both the ankles are selected to be tagged in most cases. In all the cases, the body part is selected to be tagged. In two cases, the left wrist or the left arm is selected to be tagged. Similar results are observed when taking different $\rho$ values.

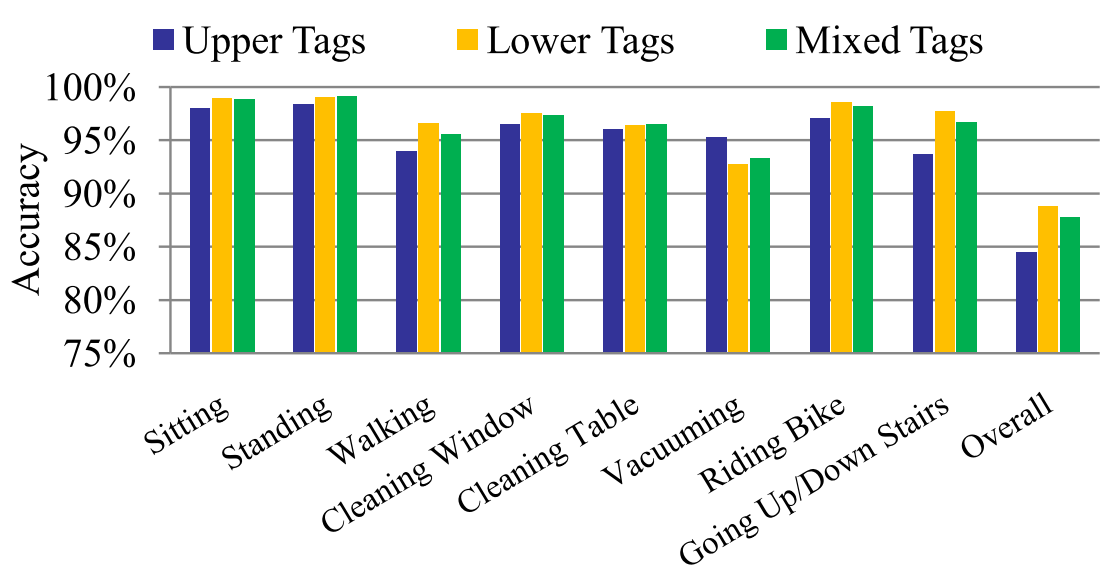

Fig. 14. Accuracy under different tag configurations.

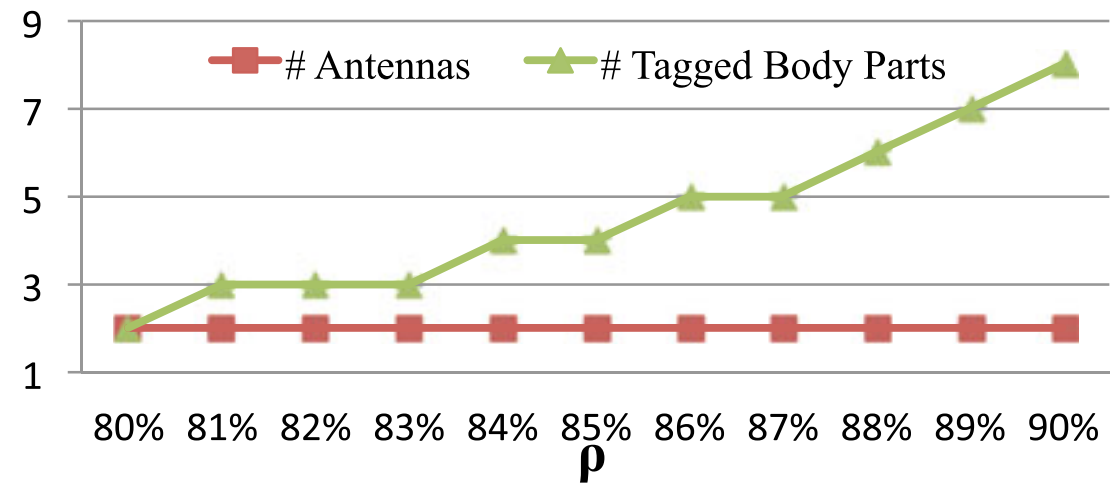

Fig. 15. The minimum number of antennas and tagged body parts required under different $\rho$ values.

Compare to the manual selection results presented in the above section, we conclude that the antennas and tags attached to the lower and upper body are effective in recognizing activities involving different lower and upper body movements, respectively. Two antennas with one placed on the upper body and the other on the lower body are sufficient to meet most accuracy requirements. A good choice is to use the mixed configuration that places the tags and antennas on one side of the upper body and the other side of the lower body. In our data set, the tags attached to the lower body are relatively more important than the tags attached to the upper body. Overall speaking, the recognition accuracy generally increases with more antennas and tags used.

### 5.4.2 Antenna Transmission Power Level

In this experiment, we evaluate the system's performance with different antenna transmission power levels (i.e., 20, 25, and 30 dbm). The existing RFID reader module allows us to set the same power level for the four antennas. Fig. 16 illustrates the overall recognition accuracies of different power levels. The recognition accuracy is above 90 percent for all transmission power levels. When the power level is set to 25 dbm, the system achieves the maximum recognition accuracy of 94 percent.

We further study the optimal transmission power levels for different subjects. As illustrated in Table 6, we discover that for all three male subjects, the optimal recognition accuracy is achieved at the power level of 25 dbm, followed by 30 and 20 dbm. This result explains the reason for the overall optimal power level of 25 dbm shown in Fig. 16. It is interesting to find that the optimal power level for the female subject is 20 dbm, followed by 25 and 30 dbm. It is possibility due to that the female subject is smaller in size and the fading effect is stronger with a lower transmission power level.

This result reveals that a higher transmission power level does not necessarily lead to a higher recognition accuracy. The optimal power level is the one most sensitive to RSS radio patterns resulted from different activities, and hence it provides the best discriminative power for activity recognition. Different subjects may have different optimal power levels, and finding the optimal power level for a subject can be easily done through training.

TABLE 5
Four Cases of Minimum Antennas and Tagged Body Parts Required when $\rho = 85\%$

| Antennas Selected | | | | Tagged Body Parts Selected | | | | | | | | |
|---|---|---|---|---|---|---|---|---|---|---|---|---|
| Back | Chest | Left Foot | Right Foot | Left Wrist | Right Wrist | Left Arm | Right Arm | Body | Left Leg | Right Leg | Left Ankle | Right Ankle |
| ✓ | | | ✓ | | | | | ✓ | | ✓ | ✓ | ✓ |
| | ✓ | ✓ | | ✓ | | | | ✓ | ✓ | | ✓ | |
| | ✓ | ✓ | | | | ✓ | | ✓ | | | ✓ | ✓ |
| | ✓ | ✓ | | | | | | ✓ | ✓ | | ✓ | ✓ |

### 5.5 Battery Consumption

In this section, we evaluate the battery consumption of our system. We measure the output current for the battery units of our reader and the recognition software system on the smartphone, respectively. For the RFID reader, the battery output current is 180, 223, and 253 mA, for the transmission power level of 20, 25, and 30 dbm, respectively. Based on this measurement, a 6,800 mAh Li battery can power the RFID reader continuously for over 24 hours without charging. During our data collection, we charged the battery during the night and collected data in the day. The system showed no sign of power shortage during the data collection.

To evaluate the battery consumption of our recognition software, we run the system on Samsung smartphone. We use a battery monitoring software built on top of the Android OS's APIs to record battery consumption. The results show our recognition software introduces an additional consumption of 38 mA. According to the measurement results presented in [45], the additional power consumption of our system on smartphone is insignificant compared to the power consumption of the OS and other applications. Also, the battery capacity of recent smartphones are typically above 2,000 mAh, an additional of 38 mA energy consumption is acceptable because a) we provide a moving solution for real-time activity recognition with a high coverage area that can support many applications; b) we achieve comparable recognition accuracy with the BSN-based solution while requiring less maintenance effort; c) our system outperforms the smartphone-based solution which only uses the smartphone's onboard sensors for activity recognition as reported in Section 5.7.

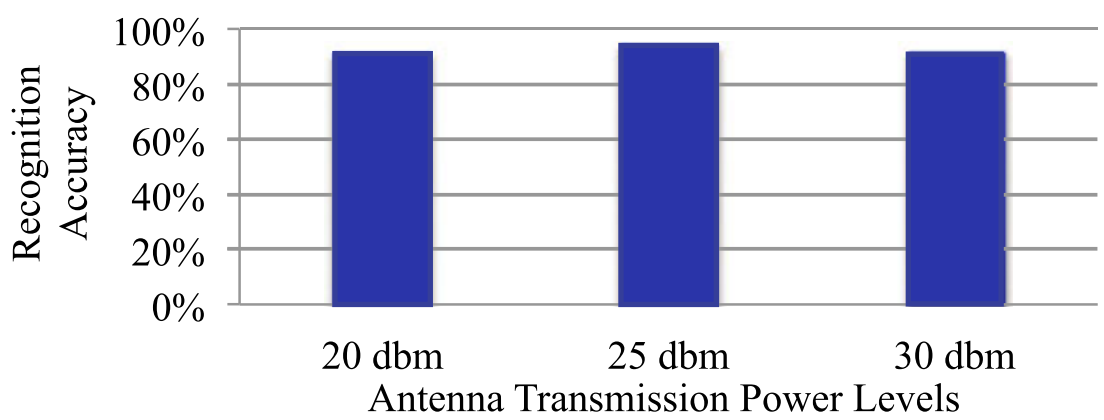

Fig. 16. Accuracy under different transmission power levels.

TABLE 6
Optimal Transmission Power Levels for different Subjects

| Subject \ PowerLevel | 20 dbm | 25 dbm | 30 dbm |
|---|---|---|---|
| 1 | 85.2% | **90.2%** | 89.1% |
| 2 | **94.6%** | 93.2% | 92.8% |
| 3 | 89.9% | **93.9%** | 91.6% |
| 4 | 84.2% | **91.0%** | 87.0% |

The above results suggest that our system is energy-efficient for performing accurate activity recognition. The current power consumption of the RFID reader is relatively high. This is because the device currently used is not optimized for our usage scenario. With the development of low-power, compact RFID readers [24], the power consumption of the RFID reader will not be a significant issue in the future. Further, as suggested in Section 5.4.2, our radio-based approach favors the most discriminative transmission power level which is often relatively lower than data-based approaches [27].

### 5.6 Effectiveness of Data Completion

We evaluate the performance of our data completion algorithm presented in Section 4.2 in this experiment. As shown in Fig. 17, by using data completion, the recognition accuracy is improved by 17.3 percent on average, suggesting that false negative reading is a major threat to the system's performance and our method effectively addresses this issue. Despite that the system's performance is relatively unstable when the sliding window size is small (i.e., 1 and 2 s), the data completion method stably increases the recognition accuracy when the window size is above 3 s. It can be seen from the figure that the data completion method significantly improves the recognition accuracy when the window size is below 10 s. For the window size above 10 s, the improvement is less significant. This result suggests that the false negative reading issue is mainly caused by not having enough time to complete reading on all four antennas, which is one of the limitations of the RFID reader we use. Our system achieves a reasonably good recognition accuracy of 80 percent without data completion when the window size is large enough (over 10 s). The false negative reading issue can be alleviated by using more sensitive readers which can read multiple antennas simultaneously, which we leave for our future work.

In summary, the data completion method is effective, and the highest recognition accuracy is achieved by using the overlap threshold of 0.7 and a sliding window size of 5 s.

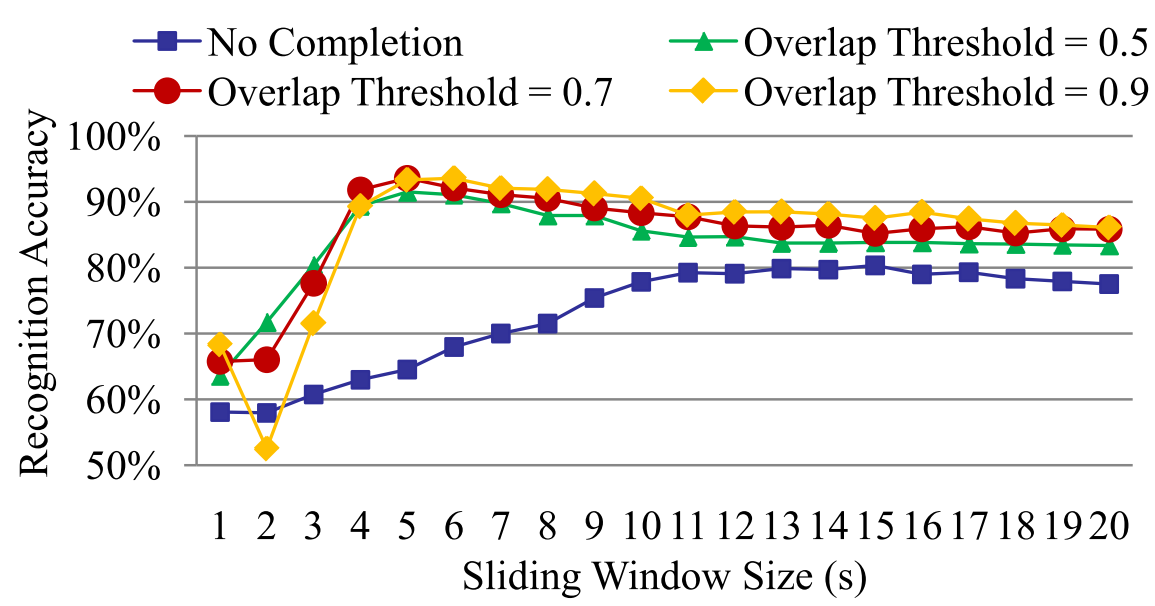

Fig. 17. Effectiveness of data completion.

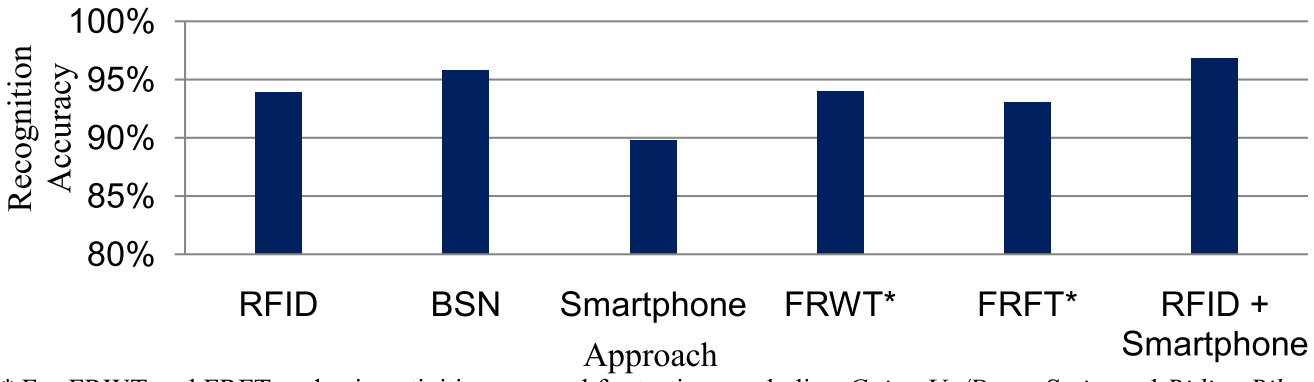

Fig. 18. Comparison between different approaches.

### 5.7 Comparison Studies

In this section, we compare the performance of the proposed approach with the BSN-based, smartphone-based approaches, and with two different setups of RFID-based systems.

In the BSN-based approach, the subject wears nine Memsic IRIS motes[5] placed at the same body parts as the RFID tags. The sampling rate of each note is 100 Hz. Data transmission is done in real-time in full power. To draw a fair comparison, we use the same data segmentation and activity recognition algorithms as used in the RFID-based system. We also extract the same temporal and spatial features from the acceleration data. For temporal features, the mean, variance, max, min, mean crossing rate, energy and entropy of the acceleration readings on each axis are extracted. For spatial features, the correlations between acceleration readings on different axes are computed.

In the smartphone-based approach, we follow the deployment strategy in the work of Kwapisz et al. [8]. A Nexus 5 smartphone is placed in the front pocket of the subject's pants. The readings from the on-board accelerometer are used as input. The same data segmentation, feature extraction, and recognition algorithms as the BSN-based approach are used.

Two alternatives of RFID-based approaches are implemented to compare with the proposed system, namely the *Fix Reader Wear Tag (FRWT)* and *Fix Reader Fix Tag (FRFT)* approaches. For the *FRWT* approach, we use the same devices as the proposed system and the subject wears the tags as described in Section 4.1. However, in this approach, we fix the RFID reader and antennas on a shelf facing the detection area. This setup is similar to RF-IDraw proposed in [15]. For the *FRFT* approach, we follow the RF-Care system proposed in [20]. The RFID reader and antennas are fixed on a shelf facing an array of tags attached to a wall. Activities are performed by a subject located between the antennas and the tags without any device attached to him. For both approaches, the RSS of tags are obtained. The same data processing pipeline as the proposed approach is used to draw a fair comparison.

For the BSN-based and smartphone-based approaches, we use all the activities shown in Table 2 for the study. For the alternative RFID-based approaches, because of their limited detection area, two activities, i.e., *Going Up/Down Stairs* which is performed in the building's staircase and *Riding Bike* which is performed outdoor, are excluded.

#### 5.7.1 Recognition Accuracy

Fig. 18 summarizes the recognition accuracies of different approaches. As shown in the figure, the recognition accuracy of both the proposed RFID- and BSN-based approaches are above 90 percent. The recognition accuracy of the BSN-based approach is 95.8 percent, which is slightly higher than the proposed approach. The slight difference in recognition accuracy is possibly caused by the randomness in the validation process. By changing the random seed used, we can also observe cases that our system slightly outperforms the BSN-based system, suggesting our solution is comparable with the BSN-based system. Most importantly, our system requires less maintenance effort than BSN. The smartphone-based approach reaches the lowest recognition accuracy of 89.8 percent (no significant change after changing the random seed). A possible explanation is that the activities carried out in our experiment involve fine-grained movements of different body parts which are difficult to be captured by a single device attached to a fixed part.

For the two alternative RFID-based approaches, the recognition accuracy of the *FRWT* and the *FRFT* approaches are 94 and 93.1 percent, respectively. However, these two approaches are tested using only six activities. For the proposed approach, the recognition accuracy for the same six activities is 94.5 percent. Considering the randomness in the validation process, this result suggests all RFID-based systems achieve comparable recognition accuracy. However, our approach outperforms the alternative approaches on detection coverage, making us applicable to more activities.

To further extend the above experiments, since our system already uses a smartphone for data collection, we combine the proposed RFID-based approach with the smartphone-based approach. Experiment result shows that by fusing RFID and smartphone sensing data, the recognition accuracy is improved to 96.8 percent, outperforming all the above approaches.

#### 5.7.2 Battery Consumption and Overall Costs

In this experiment, we compare the energy costs of the RFID reader and the BSN network for data collection. The measured output current of a single IRIS mote's battery unit is 20.3 mA. Since all the motes are identical in both hardware and firmware, the total battery consumption of the BSN network is then estimated as 182.7 mA. Compared to the measurements presented in Section 5.5, the BSN network consumes a similar amount of battery power as the RFID reader when the transmission power is set to 20 dbm. As suggested by the above results, 20 dbm is the optimal transmission power for our female subject. When set to higher transmission power levels, the current system implementation costs more power than the BSN-based approach. Because the current device used for our implementation is not optimized for wearable scenarios, battery consumption can be further reduced by using more power efficient devices.

Finally, we compare the costs of different systems. For our RFID-based solution, the RFID tags are very cheap, and the UHF RFID reader is also a low-cost COTS device widely accessible. In the BSN-based solution, we have nine IRIS motes used to build the system as presented above. While it is physically difficult to compare the detailed costs in dollars for prices vary among different distributors, the cost of our RFID reader is roughly equivalent to that of two IRIS motes. The total cost for our RFID system is less than a half of the BSN cost. Though both the RFID reader and the sensor

5. http://www.memsic.com/wireless-sensor-networks/

device are expected to drop in price in the future, the proposed solution is more cost-efficient because the BSN-based solution requires constant maintenance efforts such as charging or replacing batteries for multiple sensor devices.

## 6 Conclusion

In this paper, we present a novel wearable RFID-based system for real-time activity recognition that aims at providing an easy-to-use solution with high detection coverage to support applications like elderly care. We implement the prototype system and conduct extensive experiments using data collected in a realistic setting. The experiment results show that the proposed system can perform real-time recognition even when the delay bound is 1 s, and achieve the highest recognition accuracy of 93.6 percent when the latency is 5 s. By using the proposed antenna and tag selection algorithm, we demonstrate that a small number of properly selected antennas and tags suffice to achieve high recognition accuracy. Through comparison studies, we show that our approach achieves comparable performance with the BSN-based approach, and outperforms the smartphone-based approach.

As an initial work of using wearable RFID systems to perform activity recognition, the limitations of this work are as follows. First, the RFID system is still cumbersome to wear. To address this issue, we propose an antenna and tag selection algorithm which chooses the minimum amount of antennas and tags to achieve acceptable recognition accuracy. Further, we could use more optimized hardware solutions for mobile applications such as [24] for system implementation. Second, the number of tags can be read by a RFID reader during a single scan is limited. As a result, if multiple persons wearing our system are close to each other, the system's performance may be affected due to high false negative reading rate. This issue can partially be addressed by our antenna and tag selection algorithm which reduce the number of tags required. Moreover, since the sensing range of the RFID reader is within 3 m, the tag density can only affect the system's performance in limited cases such as multiple people embracing each other.

## Acknowledgments

This work was supported by the National 973 Project No. 2015CB352202, the NSFC under Grants 61502225, 61373011, and 61272456, and the Collaborative Innovation Center of Novel Software Technology and Industrialization.

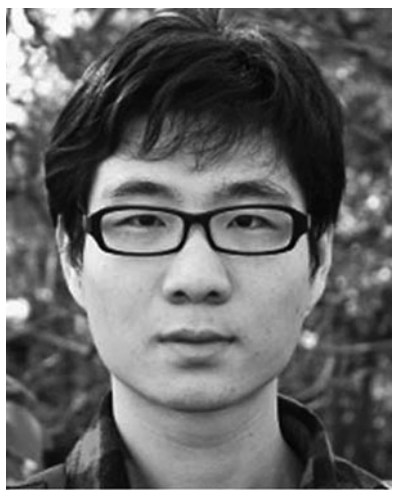

**Liang Wang** received the BSc and PhD degrees in computer science from Nanjing University in 2007 and 2014, respectively. He is currently a assistant researcher in the Department of Computer Science at Nanjing University. His research interests include pervasive and mobile computing, human computer interaction, and software engineering. He is a member of the IEEE and the ACM.

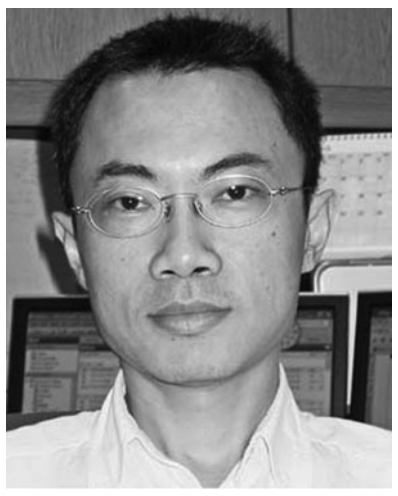

**Tao Gu** received the PhD degree in computer science from the National University of Singapore. He is currently an associate professor in the School of Computer Science and Information Technology at RMIT University, Australia. His research interests include mobile and pervasive computing, wireless sensor networks, and distributed systems. He is a senior member of the IEEE and member of the ACM.

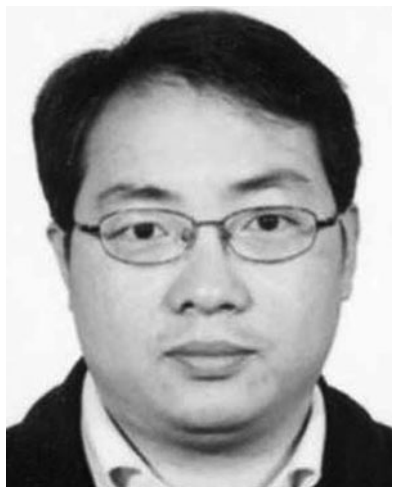

**Xianping Tao** received the PhD degree in computer science from Nanjing University, in 2001. He is currently a professor in the Department of Computer Science at Nanjing University. His research interests include software agents, middleware systems, Internetware methodology, and pervasive computing. He is a member of the IEEE.

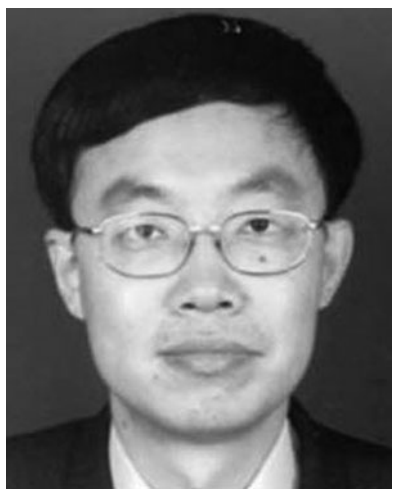

**Jian Lu** received the PhD degree in computer science from Nanjing University in 1988. He is currently a professor in the Department of Computer Science at Nanjing University. He is also the director of the State Key Laboratory for Novel Software Technology. His research interests include programming methodology, pervasive computing, software agent, and middleware. He is a member of the IEEE.